\documentclass[aps,pre,reprint,superscriptaddress,showpacs,showkeys]{revtex4-1} 
\usepackage{amsfonts}
\usepackage{amssymb}
\usepackage{amsmath}
\usepackage{graphicx}

\begin{document} 

\title{Universal properties of magnetization dynamics in polycrystalline ferromagnetic films} 

\author{Felipe~Bohn} 
\email[Electronic address: ]{felipebohn@ect.ufrn.br}
\affiliation{Escola de Ci\^{e}ncias e Tecnologia, Universidade Federal do Rio Grande do Norte, 59078-900 Natal, RN, Brazil} 
\author{Marcio~Assolin~Corr\^{e}a} 
\affiliation{Departamento de F\'{i}sica Te\'{o}rica e Experimental, Universidade Federal do Rio Grande do Norte, 59078-900 Natal, RN, Brazil} 
\author{Alexandre~Da~Cas~Viegas} 
\affiliation{Departamento de F\'{i}sica, Universidade Federal de Santa Catarina, 88010-970 Florian\'{o}polis, SC, Brazil}
 \author{Stefanos~Papanikolaou} 
\affiliation{Department of Mechanical Engineering and Materials Science and Department of Physics, Yale
University, 06520-8286 New Haven, Connecticut, USA} 
\author{Gianfranco~Durin} 
\affiliation{INRIM, strada delle Cacce 91, 10135 Torino, Italy} 
\affiliation{ISI Foundation, via Alassio 11/c, 10126 Torino, Italy} 
\author{Rubem~Luis~Sommer} 
\affiliation{Centro Brasileiro de Pesquisas F\'{i}sicas, Rua Dr.\ Xavier Sigaud 150, Urca, 22290-180 Rio de Janeiro, RJ, Brazil} 

\date{\today} 

\begin{abstract} 
We investigate the scaling behavior in the statistical properties of Barkhausen noise in ferromagnetic films. We apply the statistical treatment usually employed for bulk materials in experimental Barkhausen noise time series measured with the traditional inductive technique in polycrystalline ferromagnetic films having different thickness from $100$ up to $1000$ nm, and investigate the scaling exponents. Based on this procedure, we can group the samples in a single universality class, characterized by exponents $\tau \sim 1.5$, $\alpha \sim 2.0$, and $1/\sigma \nu z \sim \vartheta \sim 2.0$. We interpret these results in terms of theoretical models and provide experimental evidence that a well-known mean-field model for the dynamics of a ferromagnetic domain wall in three-dimensional ferromagnets can be extended for films. We identify that the films present an universal three-dimensional magnetization dynamics, governed by long-range dipolar interactions, even at the smallest thicknesses, indicating that the two-dimensional magnetic behavior commonly verified for films cannot be generalized for all thickness ranges.
\end{abstract} 



\maketitle 

\section{Introduction} 
\label{Introduction}

The magnetization dynamics and the magnetic properties of systems with reduced dimensions have been extensively studied in the recent past due to their technological relevance in a wide variety of magnetic devices. Within the several investigated effects, the Barkhausen noise (BN) appeared as one of the most usefull tools to obtain information on the dynamics of magnetic domain walls on the mesoscale~\cite{NP7p316, PRE86p066117}, and even to probe the local energetics at the nanoscale~\cite{S339p1051}.

The BN is produced by sudden and irreversible changes of magnetization, due to the irregular motion of the domain walls (DWs), as a result of the interactions with distributed pinning centers, such as defects, impurities, dislocations, and grain boundaries~\cite{Bertotti,Resumao_BN,Resumao_francesca,JMMM23p136,JMMM317p20}. In recent years, this effect has also attracted a growing interest as one of the best examples of response of a disordered system exhibiting crackling noise, whose critical dynamics exhibits random pulses (or avalanches) with scale invariance properties, power-law distributions, and universal features~\cite{N410p242,NP1p13,NP3p518,Resumao_francesca}. 

In traditional BN inductive experiments, one detects several sets of time series of voltage pulses by a sensing coil wound around a  ferromagnetic material submitted to a slow-varying magnetic field~\cite{PZ20p401, Bertotti, Resumao_BN, Resumao_francesca}. Due to its stochastic character, BN analysis usually consists in the evaluation of the statistical properties of these time series. The statistical functions are in general well described, as said, by power-laws, with critical exponents directly comparable to the predictions of theoretical models. In particular, it has been verified they exhibit universality, i.e., the exponents are independent on the microscopic details of the dynamics, being controlled only by general properties such as the system dimensionality and the range of the relevant interactions~\cite{Fractal_concepts}.  

In bulk materials, such as thin ribbons, sheets, and thick films, there is a well established and consistent interpretation of the BN statistical properties~\cite{PRL84p4705}. The traditional noise statistics includes statistical functions as the distributions of avalanche sizes and durations, the average size of an avalanche as a function of its duration, and the power spectrum, which, typically, display scaling in a quite large range, with critical exponents $\tau$, $\alpha$, $1/\sigma \nu z$, and $\vartheta$, respectively~\cite{Resumao_BN}. In the limit of vanishing magnetic field rate, these experimental exponents agree with theoretical predictions calculated for three-dimensional systems ~\cite{PRL79p4669, PRB58p6353, PRE64p066127, PRE69p026126}. Two different universality classes are found to exist, according to the range of interactions governing the DWs dynamics: (\textit{i}) \emph{long-range}, as in polycrystalline or partially crystallized amorphous alloys, due to long-range dipolar interactions, with $\tau=1.50\pm 0.05$, $\alpha=2.0\pm 0.2$ and $1/\sigma \nu z \sim \vartheta \sim2$~\cite{PRL84p4705}, and (\textit{ii}) \emph{short-range}, as in amorphous alloys, governed by short-range elastic interactions of the DW, with $\tau=1.27\pm0.03$, $\alpha=1.5\pm0.1$ and $1/\sigma \nu z\sim \vartheta \sim 1.77$~\cite{PRL84p4705}.

For ferromagnetic thin films, very important for nanostructured magnetic devices, the interpretation of the statistical properties is not so well established, due to several theoretical and experimental difficulties, even considering the recent relevant reports found in literature~\cite{PRL84p5415, IEEETM36p3090, JAP101p063903, PRL90p0872031, JAP93p6563, JMMM310p2599, NP3p547, JAP103p07D907, SSC150p1169, JAP109p07E101, PRB83p060410R}. On the theoretical side, the complexity of the magnetic domains and variety of DWs in films make the complete description of the system and the analysis of the dynamics a hard task~\cite{JSMP08020, IEEETM46p228}. On the experimental side, most of the experimental works reported so far made use of magneto-optical techniques, which restrict the analysis to avalanches sizes, and have problems in determining the correct statistical distributions as the presence of a finite observation window splits large avalanches in pieces~\cite{PRE84p061103}. Although these techniques provide a considerable gain in resolution in comparison with inductive techniques, we must note that the probed depth of the material is around $10$ nm~\cite{JAP101p063903} due to the limited penetration depth of the visible light in metals, limiting the study to the magnetic properties of the film surface. In all cases, the exponents $\tau$ measured for films are found smaller than for bulk samples. This leads to the conclusion that below $50$ nm, the magnetization dynamics is essentially two-dimensional, as expected due to reduced thickness of the studied samples~\cite{PRL84p5415, IEEETM36p3090, JAP101p063903, PRL90p0872031, JAP93p6563, JMMM310p2599, NP3p547, JAP103p07D907, SSC150p1169, JAP109p07E101, PRB83p060410R}. However, due to the restrict and insufficient amount of experimental data, the structural character and film thickness influence on the exponents is an open question and a complete comprehension of the DWs dynamics in films is still lacking.

In this paper, we report an experimental indication for an universal three-dimensional magnetization dynamics, governed by long-range dipolar interactions, in Permalloy polycrystalline ferromagnetic films down to $100$ nm, despite large variations in the macroscopic magnetic properties. We systematically investigate the scaling behavior of Barkhausen noise time series measured with the traditional inductive method. In particular, we perform a statistical analysis which includes, besides the distribution of Barkhausen avalanche sizes usually obtained for films, the distribution of avalanche durations, the average avalanche size as a function of its duration, and the power spectrum, and determine the exponents $\tau$, $\alpha$, $1/\sigma \nu z$, and $\vartheta$. Their values are all in quantitative agreement with the theoretical exponents predicted in the case of a three-dimensional dynamics of a ferromagnetic domain wall, with long-range dipolar interactions dominating the dynamics, even at the smallest thicknesses. We conclude that inductive measurements are an essential tool to investigate the magnetization dynamics in thin films, and that the usually observed two-dimensional magnetic behavior cannot be generalized for all ferromagnetic films. 

\section{Experiment} 
\label{Experiment}

For the study, we perform Barkhausen noise measurements in a set of Permalloy ferromagnetic films with nominal composition Ni$_{81}$Fe$_{19}$ and thicknesses of $100$, $150$, $200$, $500$, and $1000$ nm. The films are deposited by magnetron sputtering onto glass substrates, with dimensions $10$ mm $\times$ $4$ mm, covered with a $2$ nm thick Ta buffer layer. The deposition process is carried out with the following parameters: base vacuum of $1.5 \cdot 10^{-7}$ Torr, deposition pressure of $5.2 \cdot 10^{-3}$ Torr with a $99.99$\% pure Ar at $20$ sccm constant flow. The Ta layer is deposited using a DC source with current of $50$ mA, while the Permalloy layer is deposited using a $65$ W RF power supply. During the deposition, the substrate moves at constant speed through the plasma to improve the film uniformity. Sample thicknesses are calibrated using x-ray diffraction, which also confirms the polycrystalline character of all films. In order to induce a magnetic anisotropy and define an easy magnetization axis, a constant magnetic field of $1$ kOe is applied along the main axis of the substrate during the film deposition. Quasi-static magnetization curves are obtained with a vibrating sample magnetometer, measured along and perpendicular to the main axis of the films, in order to verify the magnetic behavior.

The Barkhausen noise time series are recorded using the inductive technique in an open magnetic circuit. This technique, although widely used to characterize bulk samples, is not commonly applied to investigate the BN in films mainly due to the low BN signal intensity. In our setup, sample and pickup coils are inserted in a long solenoid with compensation for border effects, to ensure an homogeneous applied magnetic field on the sample. The solenoid is feed by a BOP$20$-$20$ Kepco bipolar power supply/amplifier, with a low-pass filter, controlled with a DS$345$ Stanford Research Systems waveform generator. The sample is driven by a $50$ mHz triangular magnetic field, applied along the main axis, with an amplitude high enough to saturate it magnetically. BN is detected by a sensing coil ($400$ turns, $3.5$ mm long, $4.5$ mm wide, and $1.25$ MHz resonance frequency), wound around the central part of the sample. A second pickup coil, with the same cross section and number of turns, is adapted in order to compensate the signal induced by the magnetizing field. The Barkhausen signal is then amplified and filtered with a SR$560$ Stanford Research Systems low-noise preamplifier, and finally digitalized using a PCI-DAS$4020/12$ Measurement Computing analog-to-digital (A/D) converter board. All BN measurements are performed under similar experimental conditions: $100$ kHz low-pass filter set in the preamplifier and signal acquisition with sampling rate of $4$ million samples per second. The time series are acquired just around the central part of the hysteresis loop, near the coercive field, where the domain wall motion is the main magnetization mechanism~\cite{Bertotti, PRB58p6353, JMMM317p20}. 

The noise statistical analysis is performed following the procedures discussed in detail in Refs.~\cite{JMMM140p1835, F3p351, PRL84p4705, NP7p316}. Thus, a careful use of the Wiener deconvolution~\cite{NP7p316}, which optimally filters the background noise and removes distortions introduced by the response functions of the measurement apparatus in the original voltage pulses, provides us reliable statistics despite the reduced intensity of the signal. A threshold  value $v_r$ is introduced to properly define the beginning and end of each Barkhausen avalanche: the avalanche duration ($T$) is thus estimated as the time interval between these two successive intersections of the signal with $v_r$, while the avalanche size ($s$) is calculated as the integral of the signal between the same points. For each experimental run, the statistical properties are obtained from $150$ measured Barkhausen noise time series, by averaging the distributions over a $10^5 - 10^6$ avalanches. Here, a wide statistical analysis is obtained, including the distributions of Barkhausen avalanche sizes $(P(s))$ and durations $(P(T))$, the average avalanche size as a function of its duration $(\langle s(T) \rangle\, vs.\, T)$, and the power spectrum $(S(f))$. 

Having established a sophisticated method of extraction of the BN avalanches and obtained the noise statistics, the analysis of the statistical properties is done with the software \textit{BestFit}~\cite{bestFit}, which is a simple python script to perform data fitting using nonlinear least-squares minimization. The software may be applied to many multivariable problems, fitting experimental data to theory functions. We observe that the measured $P(s)$, $P(T)$, and $\langle s(T) \rangle\, vs.\, T$ avalanche distributions follow a cuttoff-limited power-law behavior and they can be, respectively, fitted as~\cite{KDahmen_unpublished}
\begin{equation}
P(s) \sim s^{-\tau}e^{-(s/s_0)^{n_s}},
\label{eq_01}
\end{equation}
\begin{equation}
P(T) \sim T^{-\alpha}e^{-(T/T_0)^{n_T}},
\label{eq_02}
\end{equation}
\begin{equation}
\langle s(T) \rangle \sim T^{1/\sigma \nu z}\left(\frac{1}{1+(T/T_0)^{n_{ave}(1/\sigma \nu z-1)}}\right)^{1/n_{ave}},
\label{eq_03}
\end{equation}
where $s_0$ and $T_0$ indicate the position where the function deviates from the power-law behavior, and $n_s$, $n_T$, and $n_{ave}$ are fitting parameters related to the shape of the cutoff function. Here, \textit{BestFit} allows us to fit them at the same time, respecting a well-known scaling relation between the exponents~\cite{PRB58p6353, PRB52p12651},
\begin{equation}
\alpha = (\tau-1)1/\sigma \nu z + 1.
\label{eq_04}
\end{equation}
We observe that the measured $S(f)$ also follows a power-law behavior at the high frequency part of the spectrum and it can be described by~\cite{JSMp01002}
\begin{equation}
S(f) \sim f^{-\vartheta}.
\label{eq_05}
\end{equation}
\noindent Although the power spectrum has not been considered for the fitting procedure, we confirm the theoretical prediction of $1/\sigma \nu z = \vartheta$, indicating that the same scaling exponent can be employed for the relation between the average avalanche size and its duration as well as for the power spectrum at high frequencies~\cite{PRB62p11699, JMMM242p1085}. In particular, it is verified that the respective scaling exponents $\tau$, $\alpha$, $1/\sigma \nu z$, and $\vartheta$ are independent of the threshold level, at least for a reasonable range of $v_r$. 

\section{Results and discussion}
\label{Results_and_discussion}

We first characterize the films from the structural and quasi-static magnetic point of view. The latter reveals a significant change in the macroscopic magnetic properties (hysteresis loops, coercive fields, etc). In contrast, the microscopic magnetization dynamics described by the scaling of Barkhausen noise appears to be universal, with a three-dimensional character, and dominated by long-range interactions.

\subsection{Structural and quasi-static magnetic characterization of the films} 

\begin{figure}[!ht] 
\includegraphics[width=8.5cm]{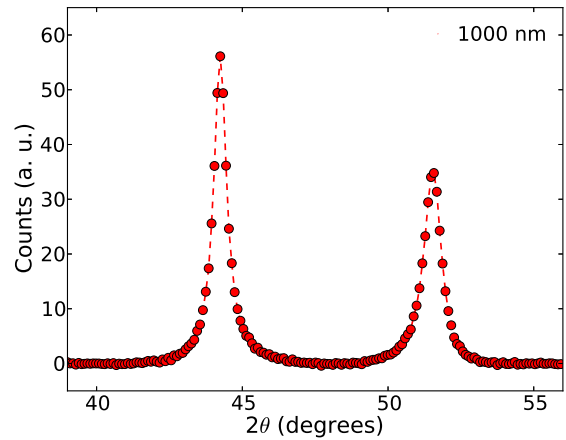} 
\vspace{-.3cm}\caption{(Color online) High angle x-ray diffraction pattern for the Permalloy film with thickness of $1000$ nm. The ($111$) and ($200$) Permalloy peaks are identified at $2\theta \sim 44.2^\circ$ and $2\theta \sim 51.5^\circ$, respectively, confirming the polycrystalline state of the film.} 
    \label{Fig_01} 
\end{figure}

While low angle x-ray diffraction is employed to calibrate the film thickness, high angle x-ray diffraction measurements are used to verify the structural character of the samples. Figure~\ref{Fig_01} shows the high angle x-ray diffraction pattern for the film with thickness of $1000$ nm. Similar results are obtained for the different thicknesses. In this case, the pattern clearly indicates the polycrystalline state, assigned by the well defined and high intensity ($111$) and ($200$) Permalloy peaks, identified at $2\theta \sim 44.2^\circ$ and $2\theta \sim 51.5^\circ$, respectively. 

\begin{figure*}[!ht] 
\hspace{-.05cm}\includegraphics[width=5.85cm]{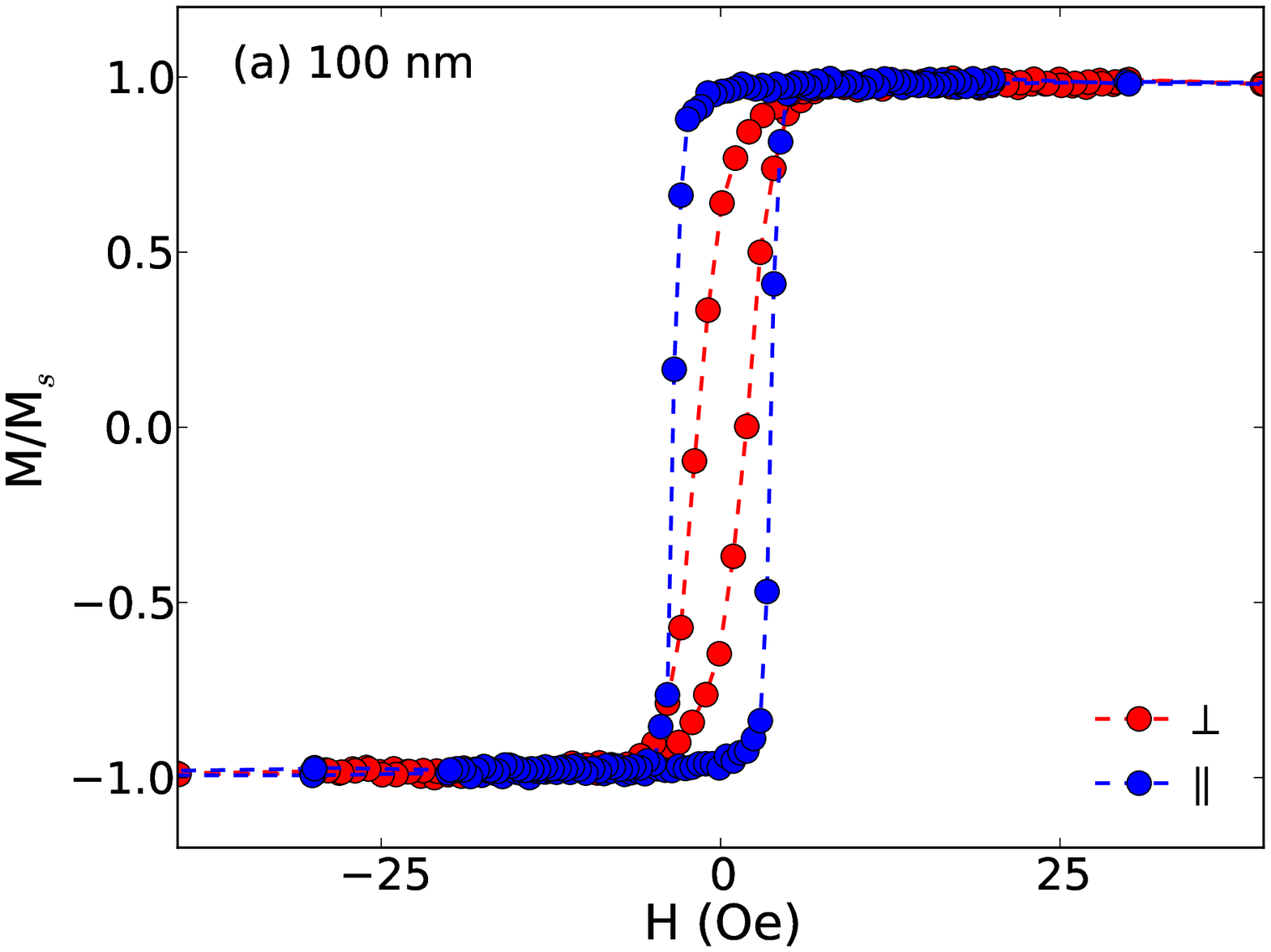} 
\hspace{-.25cm}\includegraphics[width=5.85cm]{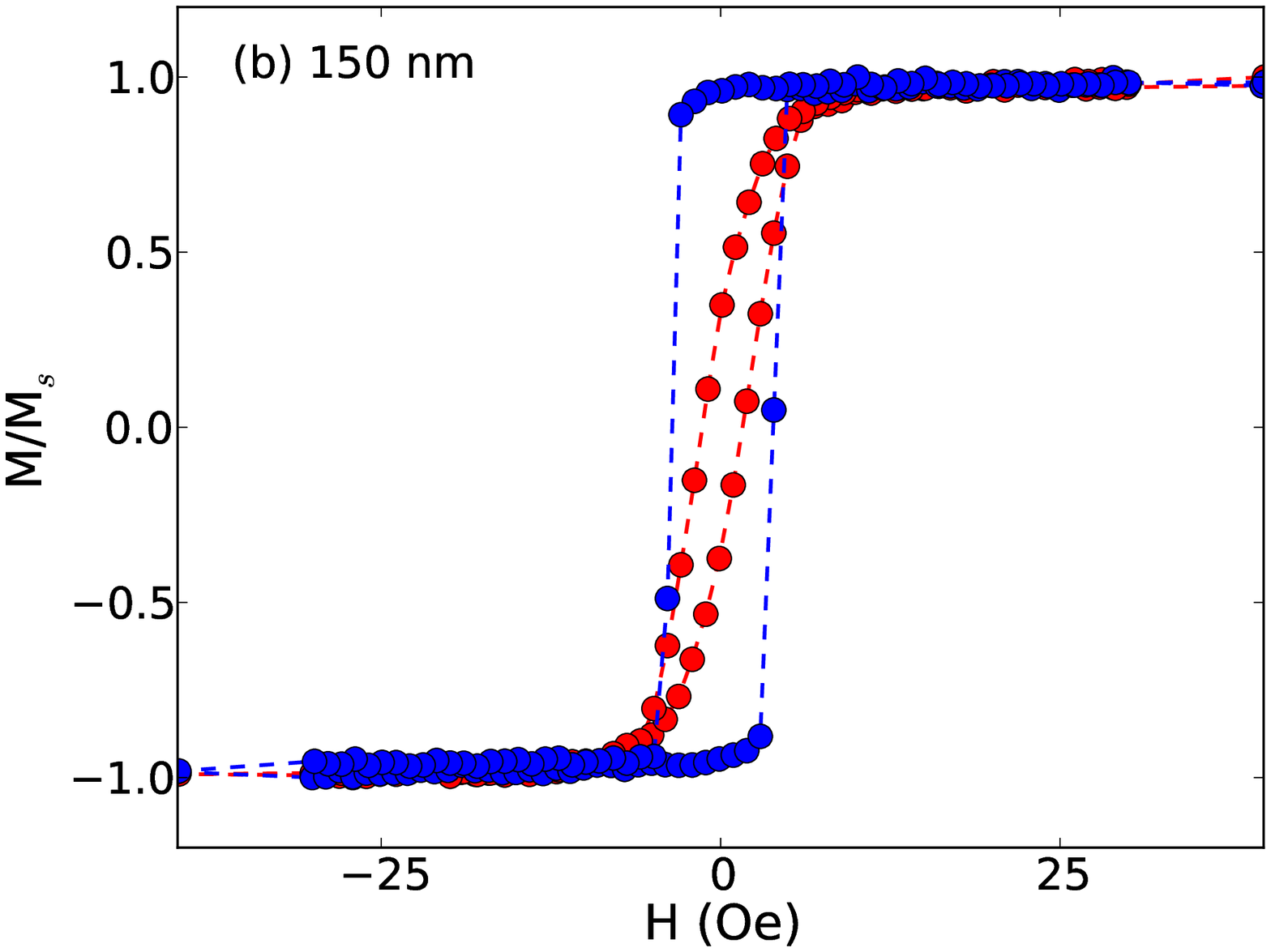}
\hspace{-.25cm}\includegraphics[width=5.85cm]{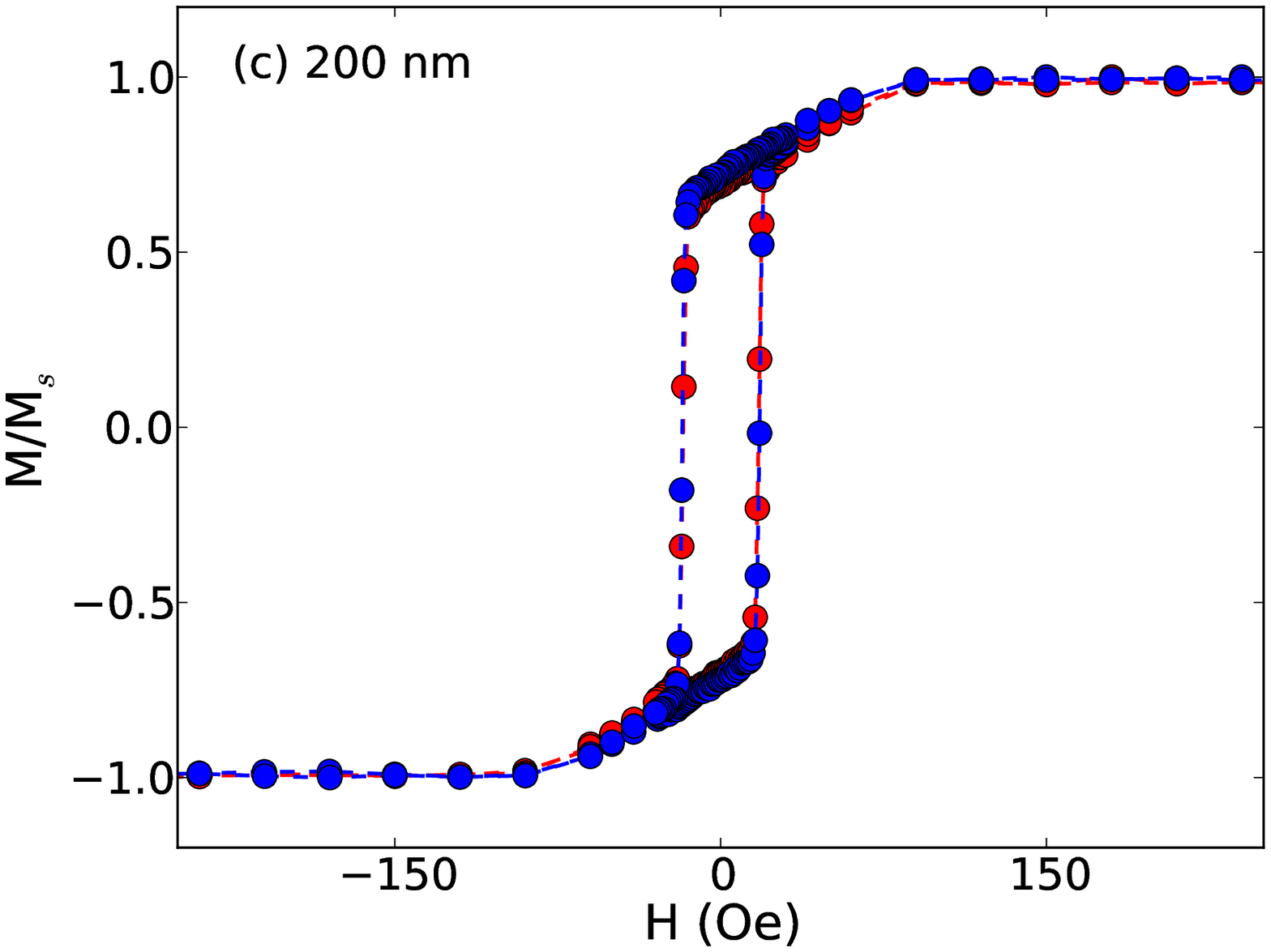} \\
\hspace{-.05cm}\includegraphics[width=5.85cm]{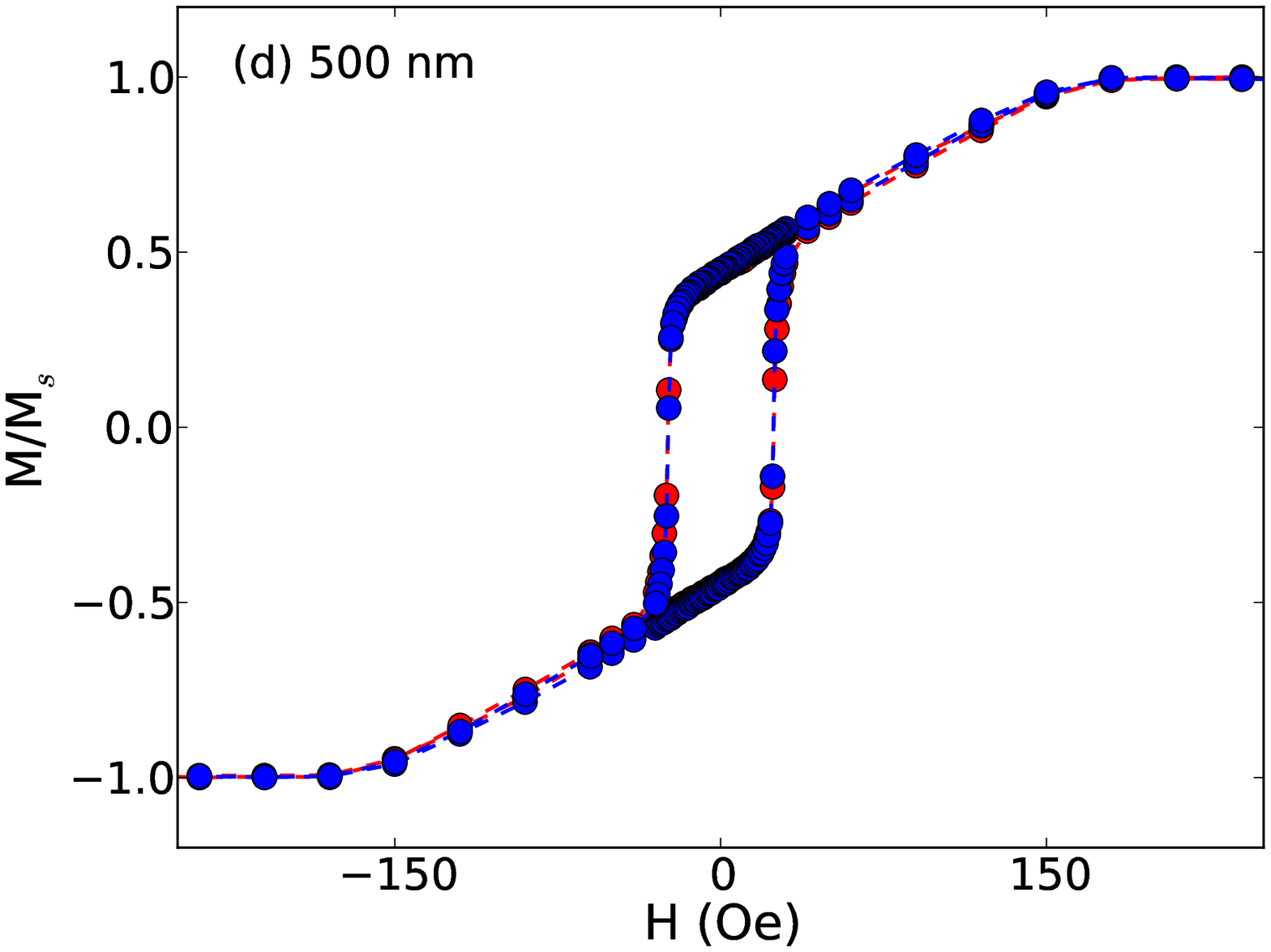}
\hspace{-.25cm}\includegraphics[width=5.85cm]{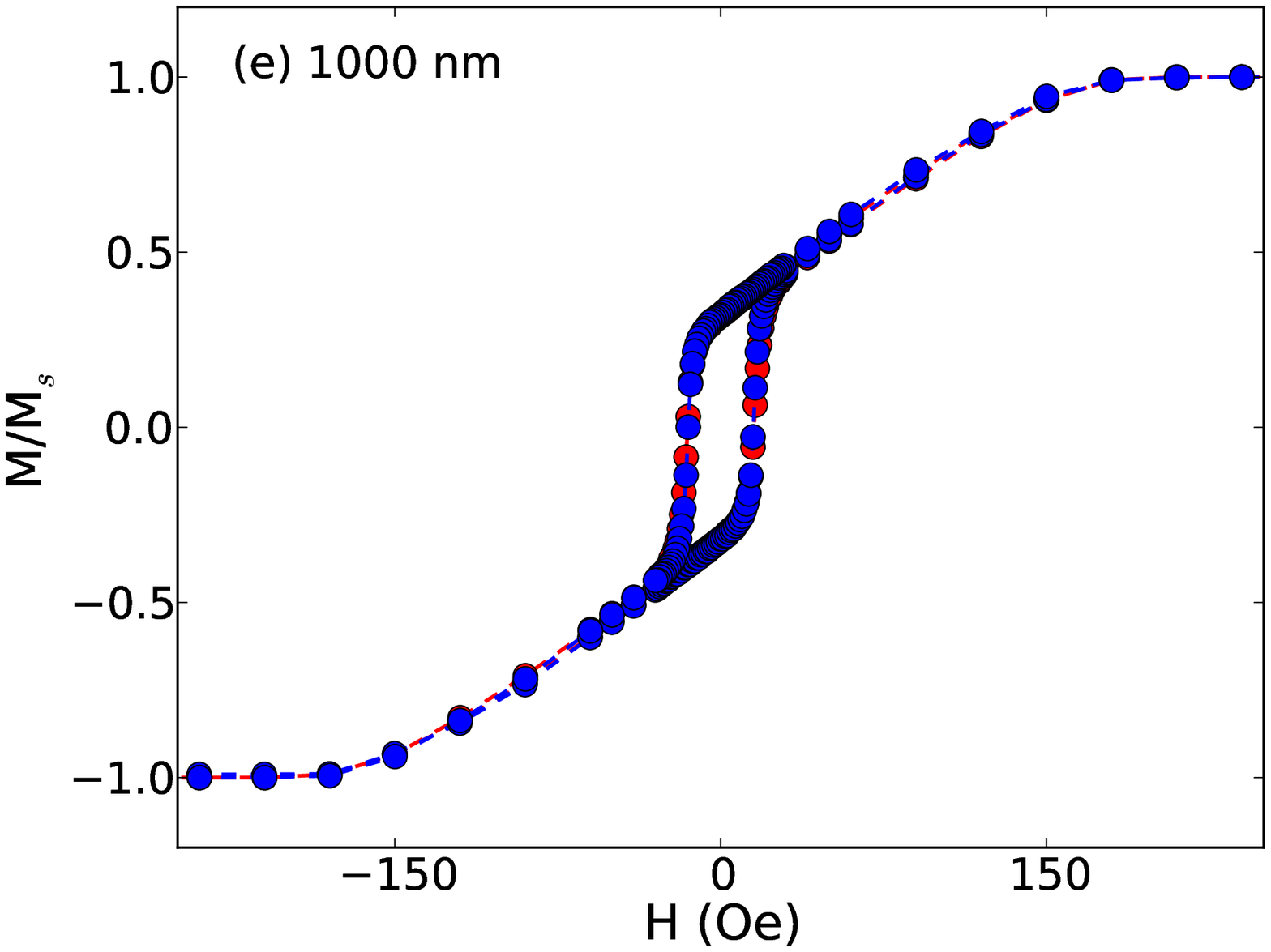} 
\hspace{-.25cm}\includegraphics[width=5.85cm]{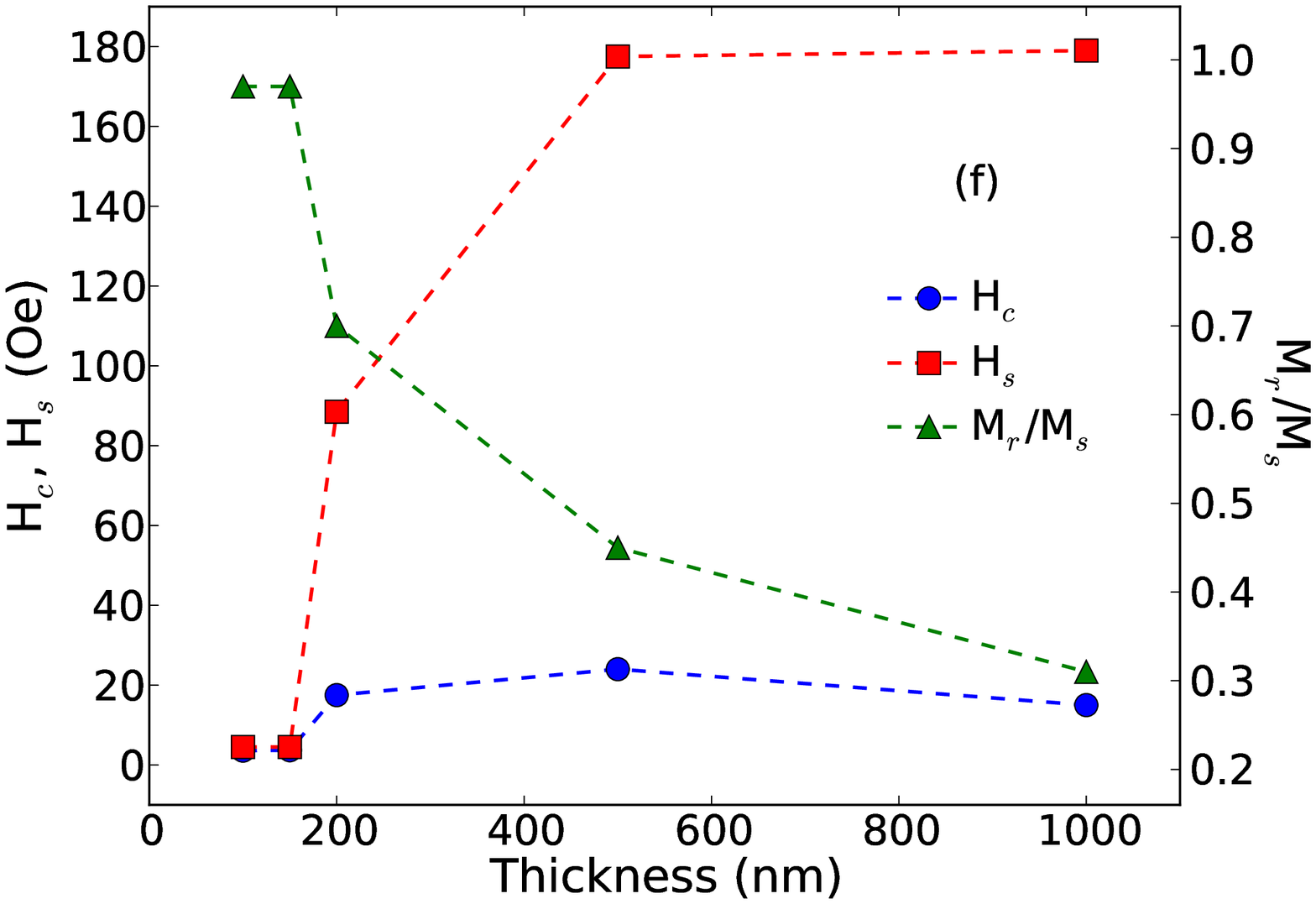} 
\vspace{-.3cm}\caption{(Color online) (a-e) Normalized quasi-static magnetization curves for the Permalloy films with different thicknesses, obtained with the in-plane magnetic field applied along ($\parallel$) and perpendicular ($\perp$) to the main axis of the films. The change of magnetic behavior is observed in the thickness range between $150$ and $200$ nm. (f) Coercive field $H_c$, saturation field $H_s$, and normalized remanent magnetization $M_r/M_s$, obtained from the magnetization curves measured along the main axis, as a function of the film thickness.} 
    \label{Fig_02} 
\end{figure*}

Magnetic characterization is obtained through magnetization curves. Figure~\ref{Fig_02}(a-e) shows the quasi-static magnetization curves, obtained with the in-plane magnetic field applied both along and perpendicular to the main axis of the films. Below $150$ nm, the angular dependence of the magnetization curves indicates an uniaxial in-plane magnetic anisotropy, induced by the magnetic field applied during the deposition process. At larger thicknesses, the curves exhibit isotropic in-plane magnetic properties, with an out-of-plane anisotropy contribution. This behavior is related to the stress stored in the film and/or to a columnar microstructure as the thickness is increased. 

The change of the magnetic behavior is usually associated to a competition between the planar demagnetization energy, the columnar shape anisotropy energy, magnetoelastic energy and the domain wall energy~\cite{JMMM282p109}. For the thinnest films, the magnetic field applied during the deposition and the planar demagnetization energy term are the main responsibles for the uniaxial and in-plane magnetic anisotropy~\cite{JMMM282p109}. For the thickest films, the in-plane anisotropy is obscured by the local stress stored in the film as the thickness increases. It is verified that the film columnar growth, due to oblique incidence of the sputtered particles, can also be responsible for creating an effective out-of-plane perpendicular magnetic anisotropy contribution~\cite{JAP87p830,IEEETM33p3634,JAP71p353, JAP31p1755}. Similar dependence of the magnetic behavior with the film thickness has been already observed and discussed in details in Refs.~\cite{Dominios_magneticos, PB384p144, JAP101p033908, JAP103p07E732, JAP104p033902}.

This evolution of the magnetization curves can also be verified through the thickness dependence of the coercive field $H_c$, saturation field $H_s$, and the normalized remanent magnetization $M_r/M_s$, obtained from the magnetization curves measured along the main axis of the films, as shown in Fig.~\ref{Fig_02}(f). The low value of $H_c$ for the thinnest films is mainly related to the uniaxial anisotropy and the existence of pinning centers for the DWs due to the surface irregularities~\cite{NIMPRB244p105}. The DWs motion is here the main magnetization mechanism, as confirmed by the almost constant values of $H_s$ and $M_r/M_s$. On the other hand, as the thickness increases, the initial increase of $H_c$ is attributed to the formation of stress centers in the bulk of the samples during the growth process~\cite{NIMPRB244p105}. The further decrease is related to the higher contribution of the magnetization rotation to magnetization process, due to the perpendicular anisotropy. This contribution is also responsible for the loss of the square shape of the magnetization curves, resulting in a considerable increase of $H_s$ and a drastic decrease of $M_r/M_s$.

\subsection{Barkhausen noise and statistical properties}

\begin{figure}[!ht] 
\includegraphics[width=8.5cm]{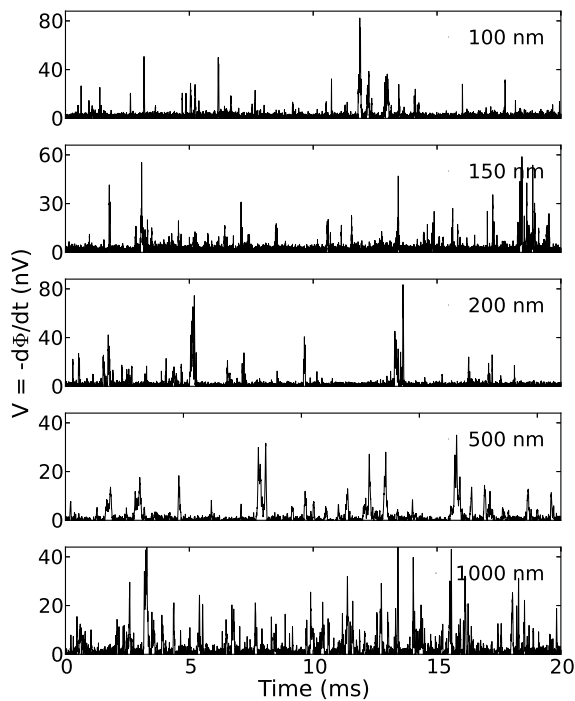} 
\vspace{-.3cm}\caption{Experimental Barkhausen noise time series measured in the Permalloy films having different thickness.} 
    \label{Fig_03} 
\end{figure}

Figure~\ref{Fig_03} shows representative examples of experimental Barkhausen noise time series measured in the Permalloy films with thicknesses between $100$ and $1000$ nm. As expected, the times series are composed of discrete and irregular avalanches, related to the sudden and irreversible changes of magnetization, a signature of the complex and jerky motion of the domain walls in the ferromagnetic films during the magnetization process. 

When compared to the BN measured in bulk materials, the visual inspection of the time series acquired in films reveals important features: the Barkhausen avalanches appear as sharper peaks, with lower amplitudes, and with shorter durations. This behavior is verified through the distributions, where the values of avalanche sizes $($from $\sim 5 \cdot 10^{-13}$ Wb up to $\sim 1 \cdot 10^{-11}$ Wb$)$ and durations $($from $\sim 1 \cdot 10^{-5}$ s up to $\sim 5 \cdot 10^{-4}$ s$)$ are some orders of magnitude lower than that obtained for bulk samples \cite{Resumao_BN}.

\begin{figure}[!h] 
    \includegraphics[width=8.5cm]{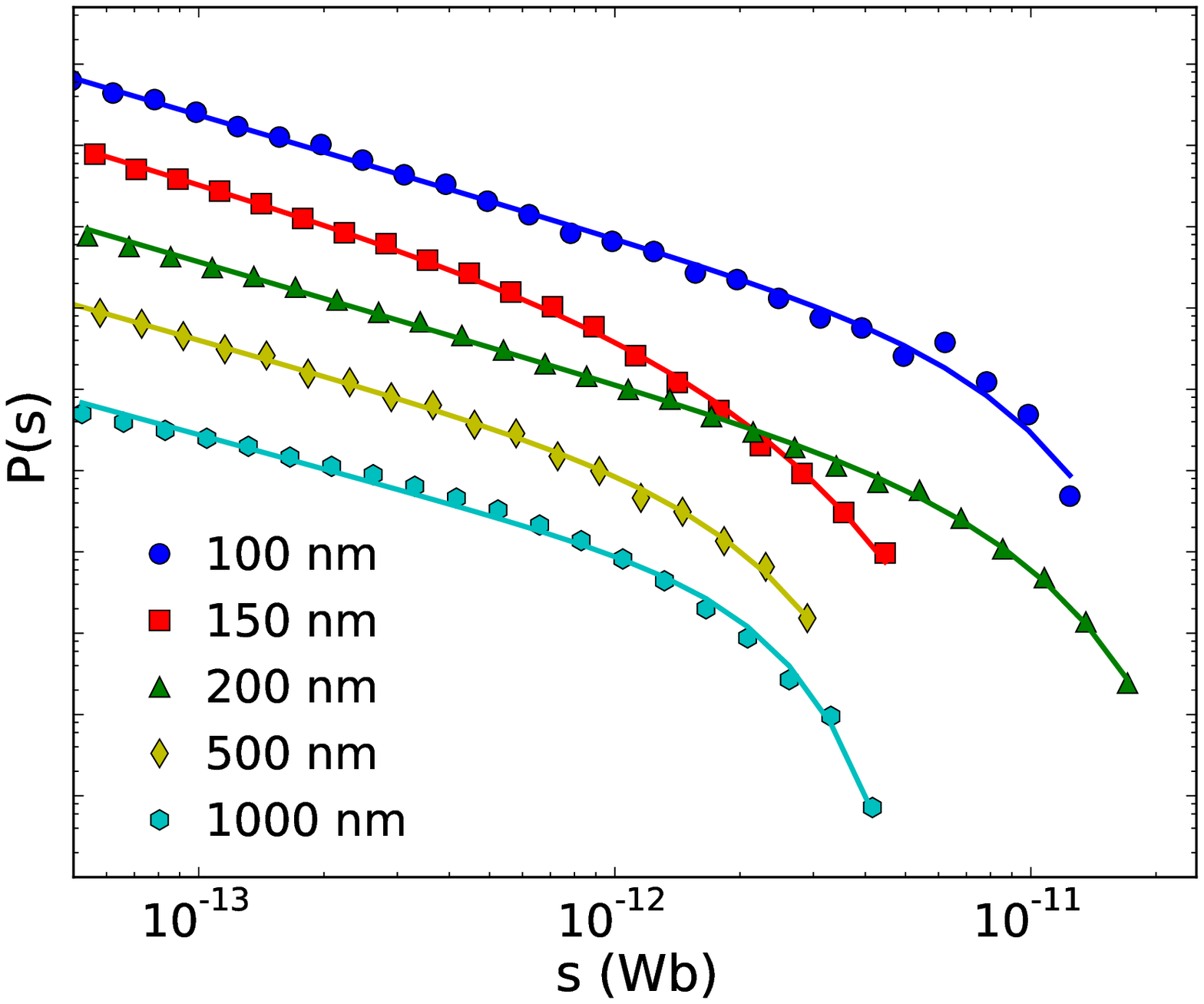} 
\vspace{-.3cm}\caption{(Color online) Distributions of Barkhausen avalanche sizes measured for the Permalloy films with different thicknesses. The solid lines are cutoff-limited power-law fittings obtained using Eq.\ (\ref{eq_01}). The best fit $\tau$ exponents are given in Table \ref{exponents}. For all Permalloy films of the set, the fittings have exponent $\tau \sim 1.5$. } 
    \label{Fig_04} 
\includegraphics[width=8.5cm]{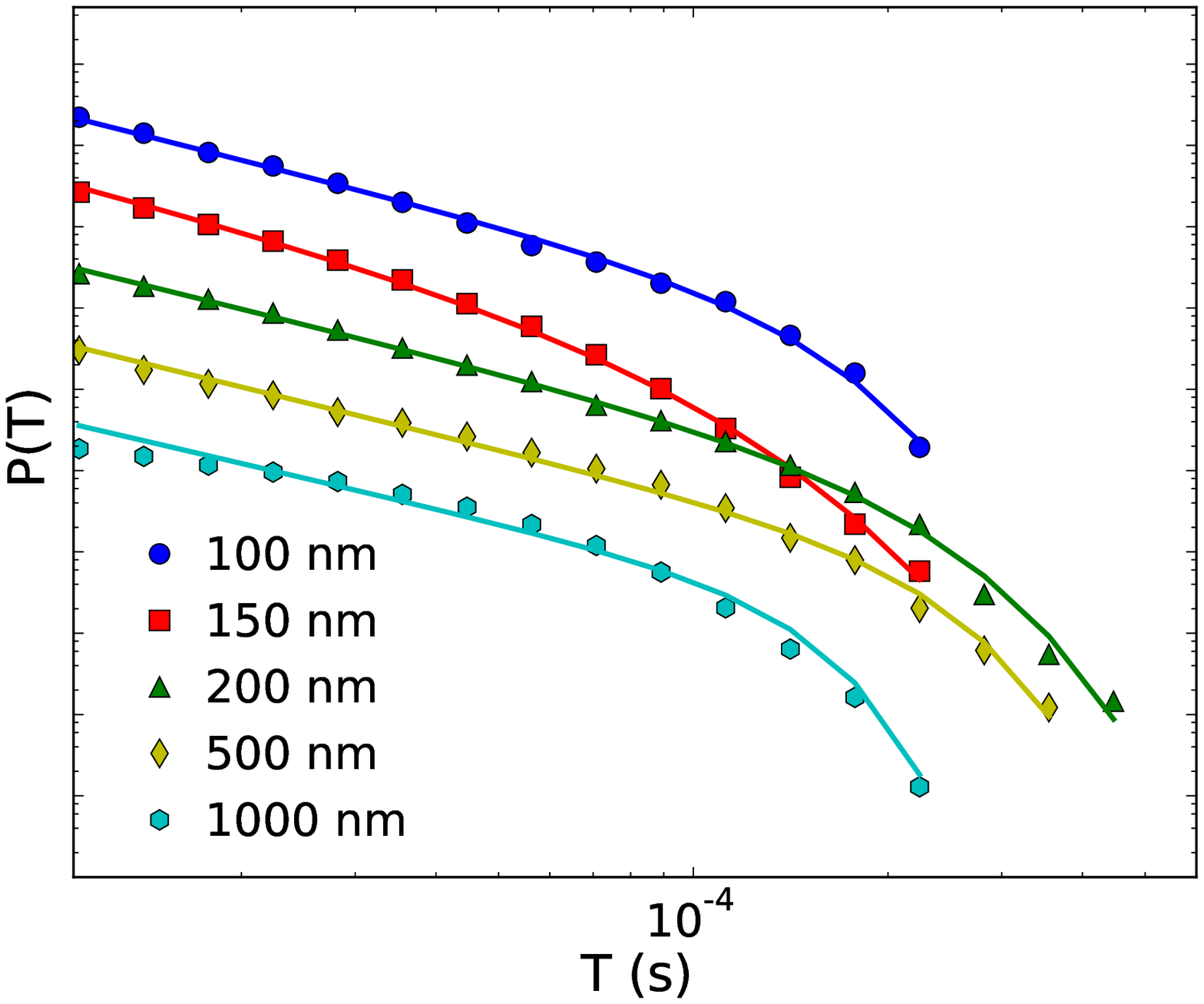} 
\vspace{-.3cm}\caption{(Color online) Distributions of Barkhausen avalanche durations measured for the Permalloy films with different thicknesses. In this case, the solid lines are cutoff-limited power-law fittings obtained using Eq.\ (\ref{eq_02}). The best fit $\alpha$ exponents are given in Table \ref{exponents}. For all Permalloy films, the fittings have exponent $\alpha \sim 2.0$.} 
    \label{Fig_05} 
\end{figure}

Figures~\ref{Fig_04} and \ref{Fig_05} show the distributions of Barkhausen avalanche sizes and durations, respectively, obtained for the Permalloy films with different thicknesses. In particular, the distributions follow a cutoff-limited power-law scaling behavior for all films and it can be fitted using Eq.\ (\ref{eq_01}) for the avalanche sizes, and Eq.\ (\ref{eq_02}) for the durations.

Thus, already from these figures, it is possible to note that the results can be grouped in a single universality class. First of all, for the distributions os avalanche sizes, the exponents are found to be $\tau \sim 1.50$, within the experimental statistical error, for all Permalloy films. Next, for the distributions of avalanche durations, the exponents are found to be $\alpha \sim 2.0$, for the very same films for all thicknesses, despite the small fluctuations due to the limited range of scaling and presence of unavoidable excess external noise at low duration for some samples with lower BN signal amplitude. 

\begin{figure}[!h] 
    \includegraphics[width=8.5cm]{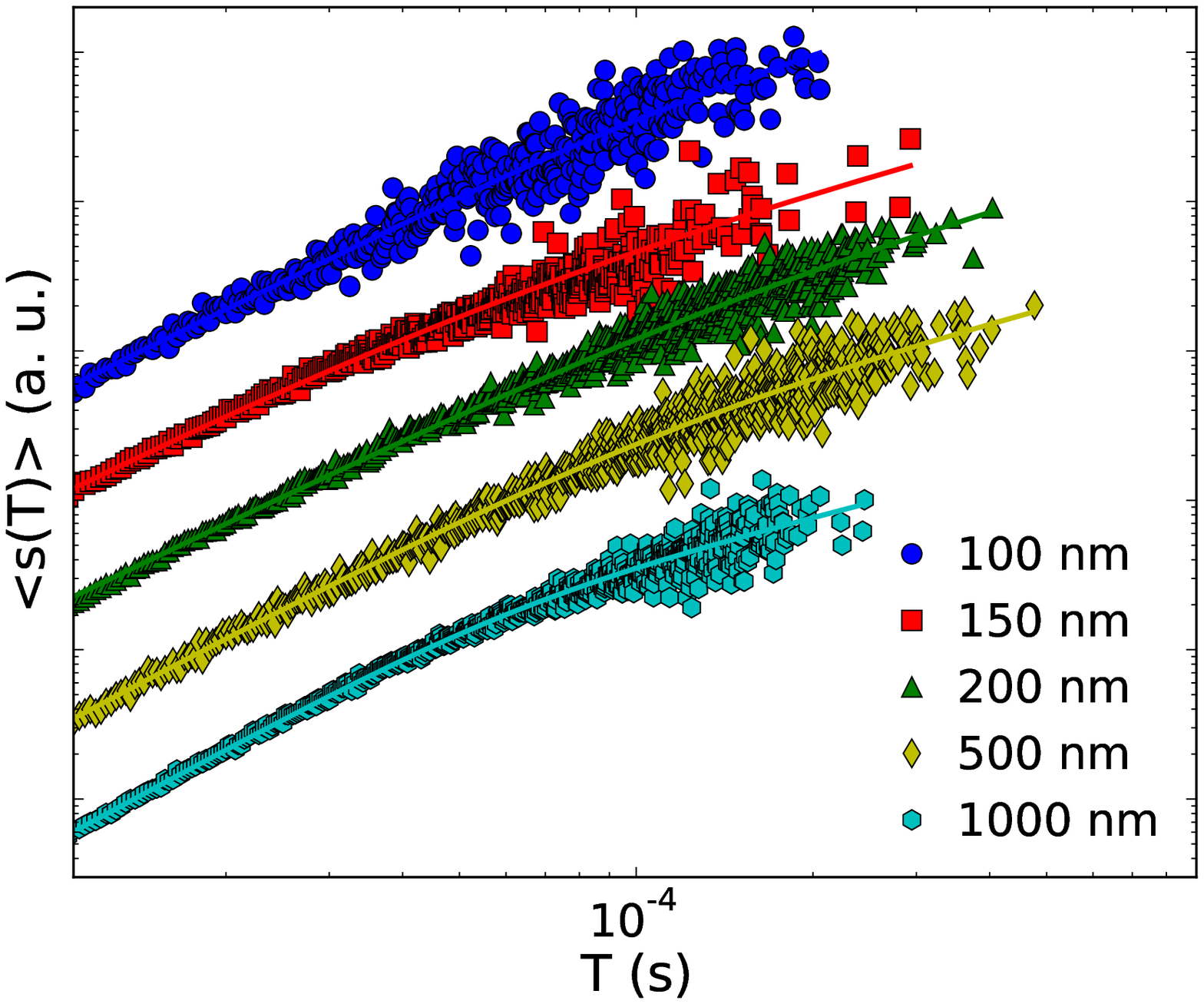} 
\vspace{-.3cm}\caption{(Color online) Average avalanche size as a function of its duration for the Permalloy films with different thicknesses. The solid lines are cutoff-limited power-law fitting obtained using Eq.\ (\ref{eq_03}). The best fit $1/\sigma \nu z$ exponents are given in Table \ref{exponents}. For all Permalloy films, the fitting have exponent $1/\sigma \nu z \sim 2.0$.} 
    \label{Fig_06} 
    \includegraphics[width=8.5cm]{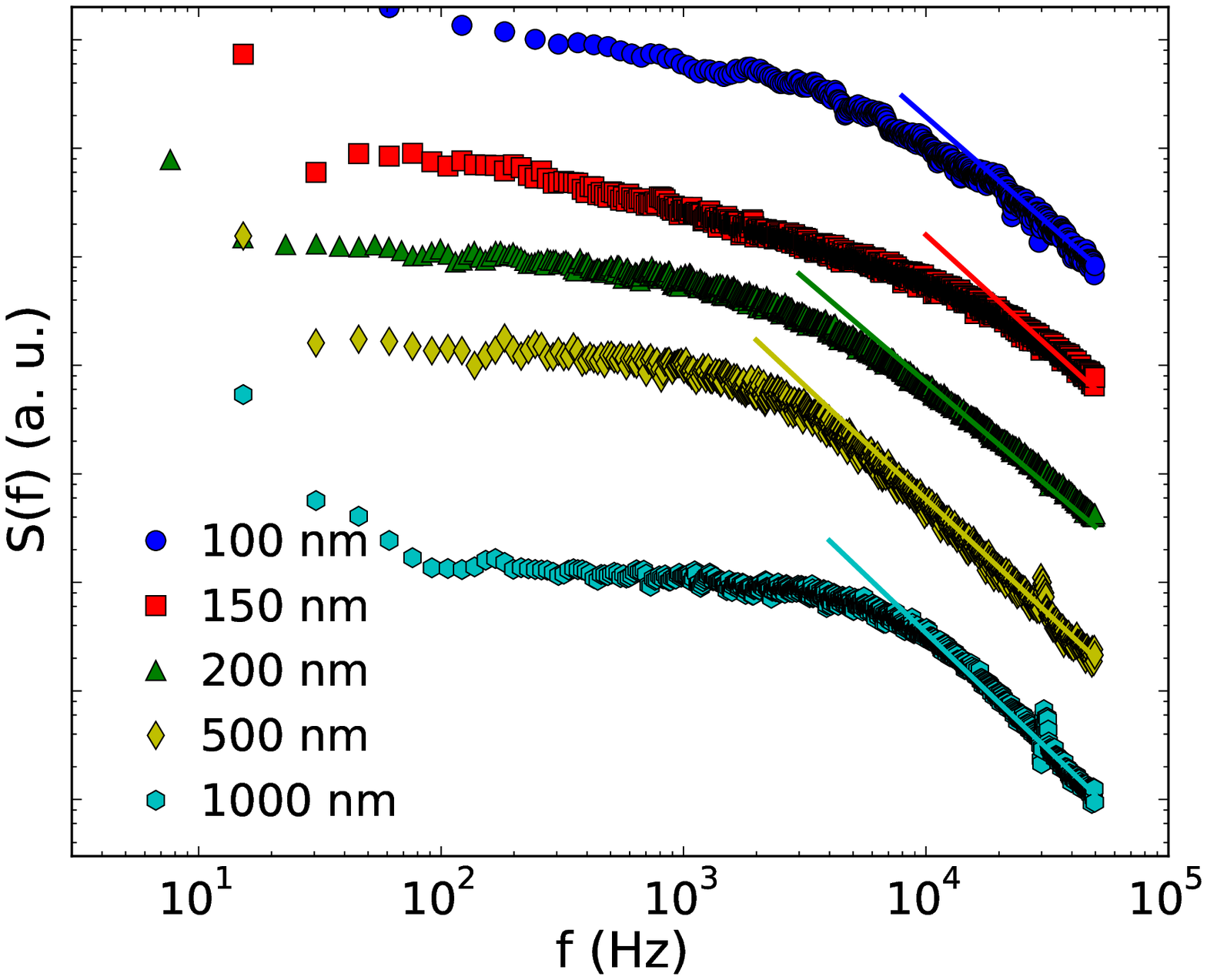} 
\vspace{-.3cm}\caption{(Color online) Power spectrum measured for the very same Permalloy films. To guide the eyes, the solid line are power-laws with slopes $\vartheta = 1/\sigma \nu z$, being $1/\sigma \nu z$ the best fit exponents given in Table \ref{exponents}. For all Permalloy films, the power-laws have exponent $\vartheta \sim 2.0$.} 
    \label{Fig_07} 
\end{figure}

When considered the average avalanche size as a function of its duration and the power spectrum, the results can be also grouped in just one universality class. Figure \ref{Fig_06} shows the curves of the average size of an avalanche as a function of its duration obtained from the very same BN times series. Similarly, this statistical function presents a cutoff-limited power-law behavior that can be fitted using Eq.\ (\ref{eq_03}). At least for the range of small durations, the power-laws are well described by exponents $1/\sigma \nu z \sim 2.0$ for all Permalloy films. The scaling relation between the average avalanche size to its duration is known as a robust quantity~\cite{JSMp01002}, becoming the most reliable test to identify the universality class of a given signal, since the exponent ${1/\sigma \nu z}$ is not influenced by problems of non-stationarity~\cite{JSMp01002} and rate effects~\cite{Resumao_BN}.

On the other side, Fig.~\ref{Fig_07} shows the power spectrum obtained for the same time series. Despite the more complex shape, a power-law behavior is observed in the high frequency part of the power spectrum and it can be described by Eq.\ (\ref{eq_05}). Although the thinner films present a restricted range of frequency where this behavior is verified, the exponent $\vartheta \sim 2.0$ seems to decribe the scaling behavior very well for all films. Remarkably, this fact is in agreement with the previously shown theoretical prediction of $1/\sigma \nu z = \vartheta$, indicating that the same scaling exponent can be employed for the relation between the average avalanche size and its duration as well as for the power spectrum at high frequencies~\cite{PRB62p11699, JMMM242p1085}. 

In this study, we estimate the scaling exponents by fitting the experimental BN statistical properties using the software \textit{Bestfit}~\cite{bestFit}. The results of the fits for the exponents are reported in Table~\ref{exponents} and also shown in Figs.~\ref{Fig_04}-\ref{Fig_06}. At a first moment, we determine the values of $\tau$, $\alpha$ and $1/\sigma \nu z$ independently by fitting $P(s)$, $P(T)$, and $\langle s(T) \rangle \, vs.\, T$ separately, as usually performed to analyse the BN statistical properties. The exponents values obtained following this procedure are close to the ones shown in the table. However, here, we determine the three critical exponents presented in Table~\ref{exponents} by jointly fitting the distributions. By fitting them at the same time, the scaling relation between the exponents, Eq.\ (\ref{eq_04}), is respected. Moreover, this scaling relation is also experimentally verified throught the agreement between the fittings and experimental results.

\begin{table}[!h]
\begin{center}
\caption{Values of $\tau$, $\alpha$ and $1/\sigma \nu z$ exponents for the experimental distributions measured for Permalloy polycrystalline ferromagnetic films with thicknesses of $100$, $150$, $200$, $500$, and $1000$ nm. The fits of $P(s)$, $P(T)$, and $\langle s(T) \rangle \, vs.\, T$ were performed simultaneously using Eqs.\ (\ref{eq_01}), (\ref{eq_02}) and (\ref{eq_03}), respecting the scaling relation between the exponents $\tau$, $\alpha$, and $1/\sigma \nu z$, Eq.\ (\ref{eq_04}).}
\label{exponents}
\begin{tabular}{cp{.25cm}cp{.25cm}cp{.25cm}c}
\hline \hline
Thickness (nm) && $\tau$ && $\alpha$ && $1/\sigma \nu z$ \\ \hline
$100$ && $1.51 \pm 0.03$ && $1.99 \pm 0.06$ && $1.95 \pm 0.04$\\
$150$ && $1.49 \pm 0.03$ && $1.99 \pm 0.06$ && $2.04 \pm 0.04$\\
$200$ && $1.49 \pm 0.02$ && $1.94 \pm 0.04$ && $1.91 \pm 0.03$\\
$500$ && $1.45 \pm 0.02$ && $1.94 \pm 0.05$ && $2.09 \pm 0.05$\\
$1000$ && $1.40 \pm 0.04$ &&  $1.85 \pm 0.07$ && $2.13 \pm 0.05$\\
\hline \hline
\end{tabular}
\end{center}
\end{table}

By considering the set of scaling exponents of the BN statistical properties in the Permalloy films, it is possible to obtain very interesting information. The first point resides in the fact that the similar values for $\tau$, $\alpha$, $1/\sigma \nu z$, and $\vartheta$ are observed for all Permalloy films, being apparently independent on the film thickness, at least at this whole range of thickness.

As the second interesting point, when compared films with distinct thicknesses, the exponents present similar values despite the expected increase of the whole sample complexity with thickness, and the strong modification of the magnetic properties observed in the magnetization curves in the thickness range between $150$ and $200$ nm. The noticeable stability of the exponents with the film thickness, altough not expected due to the strong modifications of the magnetic properties and magnetic structure with thickness, agrees with the theoretical prediction of the invariance of the exponents with respect to the increasing number of defects~\cite{PRL75p4528}, which in our case is associated to the film thickness, and corroborates the fact that the exponents are universal, i.e., independent of the microscopic details of each sample. 

As third point, the most strinking finding resides basically in the actual values of the scaling exponents. We can group the Permalloy films in an universality class characterized by exponents $\tau \sim 1.5$, $\alpha \sim 2.0$, and $1/\sigma \nu z \sim \vartheta \sim 2.0$.

The exponent $\tau$ verified here is completelly different of the ones found in several experimental works, obtained with magneto-optical techniques, for films, in particular, polycrystalline films, $\tau \sim 1.3$~\cite{PRL84p5415, IEEETM36p3090, PRL90p0872031, JMMM310p2599, JAP93p6563, NP3p547, JAP103p07D907, PRB83p060410R}. In general, the reported experimental results indicate that the exponent $\tau$ for films is smaller with respect to the ones obtained for bulk samples~\cite{PRL84p5415, IEEETM36p3090, JAP101p063903, PRL90p0872031, JAP93p6563, JMMM310p2599, NP3p547, JAP103p07D907, SSC150p1169, JAP109p07E101, PRB83p060410R}, indicating a two-dimensional magnetic behavior. However, in our case, the exponents $\tau$, $\alpha$, $1/\sigma \nu z$, and $\vartheta$ present similar values to the ones obtained for several bulk polycrystalline magnetic materials, $\tau=1.50\pm 0.05$, $\alpha=2.0\pm 0.2$ and $1/\sigma \nu z \sim \vartheta \sim2$~\cite{PRL84p4705}, suggesting a three-dimensional magnetization dynamics even at the smallest thicknesses.

To interpret the BN results with the estimation of several scaling exponents and make sure that the comparison between theoretical predictions and experimental results is meaningful, it is necessary a careful examination of the assumptions of the models found in literature~\cite{PRL79p4669, PRB58p6353, PRL84p4705, PRE69p026126, PRE64p066127, JSMP08020, IEEETM46p228, PRE69p026126, PRE64p066127}. Here, it is important to notice that the exponents measured for the Permalloy films are in quantitative agreement with the exponent values predicted by the mean-field model for the dynamics of a ferromagnetic domain wall driven by an external magnetic field through a disordered medium proposed by P.~Cizeau, S.~Zapperi, G.~Durin, and H.~E.~Stanley (CZDS model)~\cite{PRL79p4669, PRB58p6353}. CZDS model describes three-dimensional systems with a DWs dynamics governed by long-range interactions of dipolar origin and suggests a class with critical exponents of $\tau = 1.5$, $\alpha = 2.0$, and $1/\sigma \nu z =2$. More than values for the exponents, CZDS model also predicts rate-dependent exponents $P(s)\sim s^{-(3/2-c/2)}$, and $P(T)\sim T^{-(2-c)}$, where $c$ is proportional to the driving field rate, i.e., field frequency. In particular, rate-dependent $\tau$ and $\alpha$, while $1/\sigma \nu z$ and $\vartheta$ constant critical exponents have already been verified and previously reported in Ref.~\cite{NP7p316}. 

Considering the BN experimental results obtained in the Permalloy polycrystalline ferromagnetic films with different thicknesses, they not just provide experimental evidence that the CZDS model~\cite{PRL79p4669, PRB58p6353} can be extended to describe the BN statistical properties in films with three-dimensional magnetization dynamics, but the scaling exponents for the films also corroborate the universality class of polycrystalline and partially crystallized amorphous alloys, related to a DWs dynamics governed by long-range dipolar interactions, as proposed in Ref.~\cite{PRL84p4705}.

Hence, the results do not imply the existence of a new universality class in the Barkhausen noise. Since the films are thinner than polycristalline ribbons known to exhibit mean-field behavior~\cite{PRL84p4705}, but thicker than previously studied two-dimensional films~\cite{PRL84p5415, IEEETM36p3090, JAP101p063903, PRL90p0872031, JAP93p6563, JMMM310p2599, NP3p547, JAP103p07D907, SSC150p1169, JAP109p07E101, PRB83p060410R}, we interpret the results as a clear indication that the Permalloy polycrystalline ferromagnetic films, within the range of thickness between $100$ and $1000$ nm, present a typical and universal three-dimensional magnetization dynamics governed by long-range interactions of dipolar origin.

\section{Conclusion}  
\label{Conclusion}
In summary, in this paper we perform a systematic study of the Barkhausen noise statistical properties in ferromagnetic films. By applying the traditional statistical treatment employed for bulk materials, we analyse the scaling behavior in the distributions of Barkhausen avalanche sizes, distributions of avalanche durations, average size of an avalanche as a function of its duration, and power spectrum, obtained from experimental Barkhausen noise time series measured in Permalloy polycrystalline ferromagnetic films having different thickness from $100$ up to $1000$ nm. We investigate the experimental exponents $\tau$, $\alpha$, $1/\sigma \nu z$, and $\vartheta$ and compare the experimental results and theoretical predictions found in literature in order to obtain further information on the DWs dynamics in systems with reduced dimensions and understand the role of structural character and film thickness on the scaling behavior in the BN statistical properties in ferromagnetic films.

Through the results obtained with this wide statistical analysis, we can group the Permalloy films in a single universality class, characterized by exponents $\tau \sim 1.5$, $\alpha \sim 2.0$, and $1/\sigma \nu z \sim \vartheta \sim 2.0$. Thus, the exponents directly provide experimental evidence that the CZDS model~\cite{PRL79p4669, PRB58p6353} for the dynamics of a ferromagnetic domain wall driven by an external magnetic field through a disordered medium proposed can be extended for films. These results correspond to a step to understand the complex DWs motion in films and the effects of the system dimensionality and range of the interactions in the dynamics. We identify an universal three-dimensional magnetization dynamics governed by long-range dipolar interactions for the films, even at the smallest thicknesses, revealed by the quantitative agreement between the experimental results and its well-known predictions for bulk polycrystalline magnets. We emphasize that not only the $\tau$ value usually obtained when magneto-optical techniques are employed, but all the exponents $\tau$, $\alpha$, $1/\sigma \nu z$, and $\vartheta$ directly indicate that the studied films present a typical three-dimensional magnetic behavior. 

Due to the large difference between the results presented here and the ones previously reported for films, we understand that the two-dimensional magnetic behavior commonly verified for films~\cite{PRL84p5415, IEEETM36p3090, JAP101p063903, PRL90p0872031, JAP93p6563, JMMM310p2599, NP3p547, JAP103p07D907, SSC150p1169, JAP109p07E101, PRB83p060410R}, although expected due to the considered thickenesses, cannot be generalized for all thickness ranges. Here, we present part of our efforts devoted to the comprehension of the BN statistical properties in ferromagnetic films. Considering this fact, in order to obtain a complete general framework of the BN statistical properties in films in a wide range of thickness, more experimental studies are needed, not only performing a wide statistical analysis in polycrystalline and amorphous films in thinner films, but also considering studies where both inductive and magneto-optical techniques work togheter. These experiments and analyses are currently in progress.

\begin{acknowledgments} 
The research is partially supported by the Brazilian agencies CNPq (Grants No.~$310761$/$2011$-$5$, No.~$476429$/$2010$-$2$, and No.~$555620$/$2010$-$7$), CAPES, FAPERJ, and FAPERN (Grant PPP No.~$013$/$2009$, and Pronem No.~$03/2012$), Progetto Premiale MIUR-INRIM ``Nanotecnologie per la metrologia elettromagnetica'', and MIUR-PRIN 2010-11 Project2010ECA8P3 ``DyNanoMag''.
\end{acknowledgments}

\end{document}